**Critical Point of View: A Wikipedia Reader**
Editors: Geert Lovink and Nathaniel Tkacz
Editorial Assistance: Ivy Roberts, Morgan Currie
Copy-Editing: Cielo Lutino
Design: Katja van Stiphout
Cover Image: Ayumi Higuchi
Printer: Ten Klei Groep, Amsterdam
Publisher: Institute of Network Cultures, Amsterdam 2011
ISBN: 978-90-78146-13-1

**Contact**
Institute of Network Cultures
phone: +3120 5951866
fax: +3120 5951840
email: info@networkcultures.org
web: http://www.networkcultures.org

Order a copy of this book by sending an email to:
books@networkcultures.org

A pdf of this publication can be downloaded freely at:
http://www.networkcultures.org/publications

Join the Critical Point of View mailing list at:
http://www.listcultures.org

Supported by: The School for Communication and Design at the Amsterdam University of Applied Sciences (Hogeschool van Amsterdam DMCI), the Centre for Internet and Society (CIS) in Bangalore and the Kusuma Trust.

Thanks to Johanna Niesyto (University of Siegen), Nishant Shah and Sunil Abraham (CIS Bangalore) Sabine Niederer and Margreet Riphagen (INC Amsterdam) for their valuable input and editorial support. Thanks to Foundation Democracy and Media, Mondriaan Foundation and the Public Library Amsterdam (Openbare Bibliotheek Amsterdam) for supporting the CPOV events in Bangalore, Amsterdam and Leipzig. (http://networkcultures.org/wpmu/cpov/)

Special thanks to all the authors for their contributions and to Cielo Lutino, Morgan Currie and Ivy Roberts for their careful copy-editing.

# CRITICAL POINT OF VIEW
## A Wikipedia Reader

EDITED BY
**GEERT LOVINK** AND
**NATHANIEL TKACZ**
INC READER #7

# THE LIVES OF BOTS

**R. STUART GEIGER**

**Introduction: An Unlikely Candidate**

In late 2006, members of the English-language version of Wikipedia began preparing for the third annual election for the project's Arbitration Committee – or ArbCom, for short. In its own words, the dozen-or-so member committee 'exists to impose binding solutions to Wikipedia disputes that neither communal discussion, administrators, nor mediation have been able to resolve'. As they are tasked with making controversial decisions when there is no clear community consensus on a given issue, arbitrators hold some of the most powerful positions of authority in the project. In fact, ArbCom is often called Wikipedia's high or supreme court, and it should be no surprise that elections for the few seats that open each year are hotly contested. In this particular election, nominations for open seats were accepted during November 2006; according to the established rules, all editors who made at least 1,000 edits to the encyclopedia project as of October of that year were eligible to run.

In all, about 40 editors meeting these requirements nominated themselves or accepted the nominations of others, which formally involved submitting a brief statement to potential voters with reasons why they would be good arbitrators. One such candidate was an editor named AntiVandalBot, an autonomous computer program that reviewed all edits to the project as they were made and reverted those that, according to its sophisticated algorithms, were blatant acts of vandalism or spam. This bot was written and operated by a well-known administrator named Tawker, who, in a common convention, used separate user accounts to distinguish between edits he personally made and those authored by the program. AntiVandalBot's statement to voters drew on many tropes common in Wikipedian politics, including a satirical description of its accomplishments and adherence to project norms (like Neutral Point of View or NPOV) in the same rhetorical style as many other candidates:[1]

> I always express NPOV on any decision I make because I have no intelligence, I am only lines of code. I also never tire, I work 24 hours a day, 7 days a week. I think I have the most of edits of any account on this Wiki now, I have not counted since the toolserver database died. Taking a look at my talk page history, my overseers ensure that all concerns are promptly responded to. In short, a bot like me who can function as a Magic 8 Ball is exactly what we need on ArbCom! -- AntiVandalBot 05:20, 17 November 2006 (UTC)

While some Wikipedians treated the bot with at least an ironic level of seriousness, others were frustrated at Tawker, who denied he was acting through his bot and insinuated it had become self-aware. One editor removed the bot's candidate statement from the election page without prior discussion, but Tawker had AntiVandalBot quickly revert this removal of content as an act of vandalism. Another editor deleted the statement again and urged seriousness in the matter, but Tawker replaced the bot's nomination statement again, this time under his own account. Coming to the aid of his bot, Tawker passionately defended the right of any editor – human or bot – with over a thousand edits to run in the election. On cue, the bot joined in the discussion and staunchly defended its place in this political sphere by exclaiming, 'I do not like this utter bot abuse. Bots are editors too!'

I make the same argument in this chapter, although in a markedly different context. Tawker, speaking through his bot, was ironically claiming that computerized editors ought to have the same sociopolitical rights and responsibilities as human editors, capable of running for the project's highest elected position and influencing the process of encyclopedia-building at its most visible level. In contrast, I argue (with all seriousness) that these automated software agents already have a similar level of influence on how Wikipedia as a free and open encyclopedia project is constituted. However, like the elected members of ArbCom, bots are also subject to social and political pressures, and we must be careful not to fall into familiar narratives of technological determinism when asking who – or what – *actually* controls Wikipedia.

Simple statistics indicate the growing influence of algorithmic actors on the editorial process: in terms of the raw number of edits to the English-language version of Wikipedia, automated bots are 17 of the top 20 most prolific editors[2] and collectively make about 16% of all edits to the encyclopedia project.[3] On other major language versions of the project, the percentage of edits made by bots ranges from around 10% (Japanese) to 30% (French).[4] While bots were originally built to perform repetitive editorial tasks that humans were already doing, they are growing increasingly sophisticated and have moved into administrative spaces. Bots now police not only the encyclopedic nature of content contributed to articles, but also the sociality of users who participate in the community. For example, there is a policy in Wikipedia called the 'Three Revert Rule' or '3RR' that prohibits reversing another user's edits more than three times in a 24-hour period on a particular article; a bot named '3RRBot' scans for such violations and reports them to administrators. In an administrative space dedicated to identifying and banning malicious contributors (Administrative Intervention against Vandalism, or AIV), bots make about 50% of all edits, and users with semi-automated editing tools make another 30%.[5] Even bots that perform seemingly routine and uncontroversial tasks, like importing census data into articles about cities and towns, often incorporate high-level epistemic assumptions about how an encyclopedia ought to be constructed.

---

1. Note: all quotes from discussions in Wikipedia are directly copied and appear with no corrections. [sic] marks are not included due to the significant number of errors present in some of the quotes.
2. Aggregated from data collected from http://en.wikipedia.org/wiki/Wikipedia:List_of_bots_by_number_of_edits and http://en.wikipedia.org/wiki/Wikipedia:List of Wikipedians by number of edits.
3. R. Stuart Geiger, 'The Social Roles of Bots and Assisted Editing Tools', *Proceedings of the 2009 International Symposium on Wikis and Open Collaboration,* Orlando, FL: Association for Computing Machinery, 2009.
4. Felipe Ortega. 'Wikipedia: A Quantitative Analysis', Ph.D dissertation, Universidad Rey Juan Carlos, April 2009, https://www.linux-magazine.es/Readers/white_papers/wikipedia_en.pdf.
5. R. Stuart Geiger and David Ribes, 'The Work of Sustaining Order in Wikipedia: The Banning of a Vandal', *Proceedings of the 2010 Conference on Computer Supported Cooperative Work,* Savannah, GA: Association for Computing Machinery, 2010.

My goal in this chapter is to describe the complex social and technical environment in which bots exist in Wikipedia, emphasizing not only how bots produce order and enforce rules, but also how humans produce bots and negotiate rules around their operation. After giving a brief overview of how previous research into Wikipedia has tended to misconceptualize bots, I give a case study tracing the life of one such automated software agent and how it came to be integrated into Wikipedian society. HagermanBot, born 3 December 2006, now seems to be one of the most uncontroversial bots in Wikipedia, adding signatures to unsigned comments left by editors in designated discussion spaces. However, even a bot that enforced as minor of a guideline as signing one's comments generated intense debate, and the ensuing controversy reveals much detail about the dynamics between technological actors in social spaces.

**Thinking about Bots: The 'Hidden' Order of Wikipedia**
Bots have been especially neglected in existing social scientific research into the Wikipedian community. Research mentioning these computerized editors at all discusses them in one of several ways: first, as tools that researchers of Wikipedia can use for gathering sociological, behavioral, and organizational data;[6,7] second, as information quality actors (usually vandalism reversers) whose edit identification algorithms are described and effects quantitatively measured;[8,9] and third, as irrelevant entities that the software treats as humans, meaning that they must be excluded from data sets in order to get at the true contributors.[10,11,12] Researchers who have turned their attention to Wikipedia's technosocial infrastructure have discussed the significance of bots in and of themselves but make only tangential or speculative claims of their social roles.[13]

---

Most research in the third category rejects bots either for no stated rationale at all, or based on findings made in 2005 and 2006 that, at their highest levels, they only comprise about 2 to 4 percent of all edits to the site,[14] or that they are largely involved in single-use tasks such as importing public domain material.[15] As such, they have been characterized as mere force-multipliers that do not change the kinds of work that editors perform. Stivia, et al., for example, conclude their discussion of bots by describing them as one tool among others – mere social artifacts (such as standards, templates, rules, and accounts of best practices) that are 'continually created to promote consistency in the content, structure, and presentation of articles'.[16] Their discussion of information quality, like most discussions of Wikipedia, is focused on the actions of human editors. In such a view, bots do not perform normative enforcement of standards. Rather, 'power editors' use bots – along with rules and templates – in the same way that a police officer uses a car, ticket book, legal code, and a radar gun to perform a more efficient and standardized form of normative enforcement. While the authors do reveal important aspects of Wikipedia's infrastructures, they are largely focused on unraveling the complicated standards and practices by which editors coordinate and negotiate. Research into Wikipedia's 'policy environment'[17] or various designated discussion spaces has operated on this same human-centered principle, demonstrating the complex and often 'bureaucratic'[18] procedures necessary for the project's functioning.

Most interesting is that bots are invisible not only in scholarship, but in Wikipedia as well; when a user account is flagged as a bot, all edits made by that user disappear from lists of recent changes so that editors do not review them. Operators of bots have also expressed frustration when their bots become naturalized, that is, when users assume that the bot's actions are features of the project's software instead of work performed by their diligent computerized workers. In general, bots tend to be taken for granted, and when they are discussed, they are not largely differentiated from human editors. As with any infrastructure, technological artifacts in Wikipedia have generally been passed over, even as they have been

---

incorporated into everyday yet essential maintenance activities. While such a view may have been appropriate when it was first made – around 2004 and 2005 – significant developments in bot operation have resulted in a massive increase in the number and scope of bot edits. Despite this, recent research into the project largely passes over bots, operating under the assumption that the role of such technological actors has not changed.

**Articulations of Delegation**
Taking from sociologist of science and technology Bruno Latour's famous example, I argue that bots are not mere tools but are instead closer to the speed bumps he analyzes as social actors. While Latour, along with other actor-network theorists, defends a functional equivalence between human and non-human actors in their ability to engage in social activities, he stresses that the nature of the task being performed and the constellation of actors around it can be fundamentally changed when delegated to a technological actor instead of a human one. As Latour describes, a neighborhood that decides to punish speeding cars can delegate this responsibility to police officers or speed bumps, which seem to perform roughly equivalent actions. Yet compared to police officers, speed bumps are unceasing in their enforcement of this social norm, equally punishing reckless teenagers and on-call ambulances.

As Latour argues, the speed bump may appear to be 'nonnegotiable', [19] but we must not be fooled into thinking that we have 'abandoned meaningful human relations and abruptly entered a world of brute material relations'. [20] Instead, he insists that we view technologies as interdependent social actors and trace the network of associations in which they operate. Within this broader view, it may actually be easier to negotiate with speed bumps than a police officer, particularly if a city's public works department is more open to outside influence than the police department. As such, Latour rejects the distinction between matter and discourse when analyzing technologies in society, arguing that 'for the engineers, the speed bump is one *meaningful articulation* within a gamut of propositions'. [21] This methodology demands that we trace the ways in which actors articulate meaning, with the critical insight that both the actors and the articulations can (and indeed, must) be either human or non-human:

> In artifacts and technologies we do not find the efficiency and stubbornness of matter, imprinting chains of cause and effect onto malleable humans. The speed bump is ultimately *not* made of matter; it is full of engineers and chancellors and lawmakers, commingling their will and their story lines with those of gravel, concrete, paint, and standard calculations. [22]

Similar to Latour's speed bumps, Wikipedian bots are non-human actors who have been constructed by humans and delegated the highly social task of enforcing order in society. Bots also appear to be as non-negotiable as speed bumps, with their creators seemingly able to dominate the unsuspecting masses with their technical skills and literally remake Wikipedia in their own image. We must pay close attention to both the material and semiotic conditions in which bots emerge within the complex collective of editors, administrators, committees, discussions, procedures, policies, and shared understandings that make up the social world of Wikipedia. Following Latour, we gain a radically different understanding of bot operations if we trace out how a collective articulates itself, and particularly if we pay attention to the different ways they are 'commingling their will and their story lines' to other humans and non-humans. Bots, like infrastructures in general, [23] simultaneously produce and rely upon a particular vision of how the world is and ought to be, a regime of delegation that often sinks into the background – that is, until they do not perform as expected and generate intense controversies. In these moments of sociotechnical breakdown, these worldviews are articulated in both material and semiotic modes, and are rarely reconciled by either purely technological or discursive means.

---

19. Bruno Latour, *Pandora's Hope: Essays on the Reality of Science Studies*, Cambridge, Mass: Harvard University Press, 1999, p. 187.
20. Ibid.
21. Ibid.
22. Ibid, p. 190.

These aspects of bots in Wikipedia are best illustrated by the story of HagermanBot, programmed with the seemingly uncontroversial task of appending signatures to comments in discussion spaces for those who had 'forgotten' to leave them. While the discursive norm to sign one's comments had been in place for some time – with human editors regularly, but not universally, leaving replacement signatures – a growing number of editors began to take issue with the bot's actions. This controversy illustrated that a particular kind of normative enforcement and correction, while acceptable when casually performed on a fraction of violations sometimes days or weeks after, became quite different when universally and immediately implemented by a bot. As Wikipedians debated the issue, it became clear that the issue concerned far more than whether people ought to sign their comments. High-level issues of rights and responsibilities began to emerge, and the compromise, which I argue has served as the basis for relations between human and robotic editors, was manifested at a technical level as an opt-out mechanism. However, this technical compromise was undergirded by the social understanding that 'bots ought to be better behaved than people', as one administrator expressed it – and both aspects of this resolution still undergird bot development in Wikipedia to this day.

**Case Study: HagermanBot, A Problem and a Solution**
Wikipedians conduct a significant amount of communication through the wiki, and designated discussion (or talk) spaces are, at the software level, functionally identical to the collaboratively-edited encyclopedia articles. To add a comment, a user edits the discussion page, appends a comment, and saves the new revision. Unlike the vast majority of online communication platforms, such as message boards, chat rooms, or email listservs, the wiki is not specifically designed for communication and thus functions quite differently. For example, malicious users can remove or edit someone else's comments just as easily as they can edit an encyclopedia article – although this is highly discouraged and moderated by the fact that the wiki platform saves a public history of each revision. In 2006, a user called ZeroOne noted another problem arising in discussion spaces: many Wikipedians made comments without leaving a signature, making it difficult to determine not only who made a certain statement, but also when it was made. A user could go through the revision histories to find this information, but it is tedious, especially in large discussions. However, as with many tedious tasks in Wikipedia, a few editors sensed that there was a need for someone to do this work – users like ZeroOne.

---

23. Susan Leigh Star, 'The Ethnography of Infrastructure', American Behavioral Scientist 43:3 (November 1999): 377-391.

At 06:15 on 17 October 2006, user ZeroOne made his 4,072nd contribution to Wikipedia, editing the discussion page for the article on 'Sonic weaponry'. Instead of adding a comment of his own about the article, he merely appended the text {{unsigned|71.114.163.227|17 October 2006}} to the end of a comment made by another user about twenty-five minutes earlier [05:50]. When ZeroOne clicked the submit button, the wiki software transformed his answer into a pre-formatted message. Together, the edits of 71.114.163.227 and ZeroOne added the following text to the article's designated discussion page:

> Ultrasound as a weapon is being used against American citizens in Indiana. Any experts out there wish to make a study, look to Terre Haute, maybe its the communication towers, that is my guess. It is an open secret along with its corrupt mental health system. – Preceding unsigned comment added by 71.114.163.227 (talk · contribs) 17 October 2006

Two minutes later [06:17], ZeroOne performed the same task for an unsigned comment made by a registered user on the talk page for the 'Pseudocode' article – adding {{unsigned|BIueyoshi321|17 October 2006}}. About two hours later [08:40], he spent twenty minutes leaving {{unsigned}} messages on the end of eight comments, each made on a different discussion page. While ZeroOne could have manually added the text to issue the message, this process was made standard and swift because of templates, a software feature that enables users to issue pre-formed messages using shorthand codes.

While the existence of templates made ZeroOne's work somewhat automated, this editor felt that it could be made even more so with a bot. ZeroOne soon posted this suggestion in a discussion space dedicated to requests for new bots. Over the next few weeks, a few users mused about its technical feasibility and potential effects without making any concrete decisions on the matter. The discussion stagnated after about a dozen comments and was automatically moved into an archive by a bot named Werdnabot on 16 November 2006, after having been on the discussion page for fourteen days without a new comment. Yet in the next month, another user named Hagerman was hard at work realizing ZeroOne's vision of a bot that would monitor talk pages for unsigned comments and append the {{unsigned}} template message without the need for human intervention, although it is unclear if Hagerman knew of ZeroOne's request. Like ZeroOne, Hagerman had used the template to sign many unsigned comments, although many of these were his own comments instead of ones left by others.

On 30 November 2006, having finished programming the bot, Hagerman registered a new user account for HagermanBot and wrote up a proposal the next day. In line with Wikipedia's rules on bot operation, Hagerman submitted his proposal to the members of the Bot Approval Group (BAG), an ad-hoc committee tasked with reviewing bot proposals and en-suring that bots are operated in accordance with Wikipedia's policies. Tawker, the operator of AntiVandalBot and a member of the BAG, asked Hagerman for a proof of concept and asked a technical question about how the bot was gathering data. Hagerman provided this information, and Tawker approved the bot about 24 hours later, with no other editors taking part in the discussion. On 00:06 on 3 December, it began operation, automatically appending specialized {{unsigned}} messages to every comment that it identified as lacking a signature. The first day, 790 comments were autosigned, and HagermanBot made slightly over 5000 edits over the next five days. By the end of December 2006, HagermanBot had become one of the most prolific users to edit Wikipedia in that month, outpacing all other humans and almost all other bots.

**A Problem with the Solution**
There were a few problems with the bot's identification algorithms, making it malfunction in certain areas: programming errors that Hagerman promptly fixed. However, some users were annoyed with the bot's normal functioning, complaining that it instantly signed their comments instead of giving them time to sign their own comments after the fact. For these editors, HagermanBot's message was 'embarrassing', as one editor stated, making them appear as if they had blatantly violated the Signatures guideline. Others did not want bots editing messages other users left for them on their own user talk pages as a matter of principle, and an equally vocal group did not want the bots adding signatures to their own comments.

While Hagerman placated those who did not want the bot editing comments left for them, the issue raised by the other group of objecting editors was more complicated. These users were, for various reasons, firmly opposed to having the bot transform their own comments. One user in particular, Sensemaker, did not follow what was claimed to be the generally-accepted practice of using four tildes (~~~~) to automatically attach a linked signature and timestamp, instead manually adding '-Sensemaker' to comments. HagermanBot did not recognize this as a valid signature and would therefore add the {{unsigned}} template message to the end, which Sensemaker would usually remove. After this occurred about a dozen times in the first few days of HagermanBot's existence, Sensemaker left a message on Hagerman's user talk page, writing:

> HangermanBot keeps adding my signature when I have not signed with the normal four tilde signs. I usually just sign by typing my username and I prefer it that way. However, this Bot keeps appearing and adding another signature. I find that annoying. How do I make it stop? -Sensemaker

Like with the previous request, Hagerman initially responded quickly, agreeing to exclude Sensemaker within ten minutes of his message and altering the bot's code fifteen minutes later. However, Hagerman soon reversed his position on the matter after another editor said that granting Sensemaker's request for exclusion would go against the purpose of the bot, emphasizing the importance of timestamps in discussion pages. Sensemaker's manual signature did not make it easy for a user to see when each comment was made, which Fyslee, a vocal supporter of the bot, argued was counterproductive to the role of discussion spaces. Hagerman struck the earlier comments and recompiled the bot to automatically sign Sense-maker's comments, again calling Fyslee's remarks 'Very insightful!' As may be expected, Sensemaker expressed frustration at Hagerman's reversal and Fyslee's comment – in an unsigned comment which was promptly 'corrected' by HagermanBot.

Yet for Sensemaker and other editors, it was not clear 'who gave you [Hagerman] the right to do this', as one anonymous user who contested HagermanBot exclaimed. Hagerman responded to such rights-based arguments by linking to his bot proposal, which had been approved by the Bot Approval Group – clearly able to enroll this committee as an ally in defense of the bot. In fact, it seemed that Hagerman had a strong set of allies: a growing number of enthusiastic supporters, the BAG, the Signatures guideline, ideals of openness and transparency, visions of an ideal discursive space, the {{unsigned}} template, and a belief that signing unsigned comments was a routine act that had long been performed by humans. Yet for some reason, a growing number of editors objected to this typical, uncontroversial practice when HagermanBot performed it.

Many users who had previously left their comments unsigned or signed with non-standard signatures began to make themselves visible, showing up at Hagerman's user talk page and other spaces to contest what they portrayed as an unfair imposition of what they believed ought to be optional guidelines. The anti-HagermanBot group was diverse in their stated rationales and suggested solutions, but all objected to the bot's operation on some level. Some objectors staunchly opposed any user signing their comments, bot or human, and took issue with the injunction to sign one's comments using the four tilde mechanism – Sensemaker was one of these editors, although others did not want to use a signature at all. Another group did not want to see a bot universally enforcing such a norm, independent of their stance on the necessity of signatures:

> I don't really like this bot editing people's messages on other people's talk pages without either of their consent or even knowledge. I think it's a great concept, but it should be an opt-in thing (instead of opt-out), where people specify with a template on their userpage if they want it, like Werdnabot, it shouldn't just do it to everyone. Just my two cents. --Rory096 01:36, 11 December 2006 (UTC)

Having failed to convince Hagerman, Sensemaker shifted venues and brought the issue to the members of the Bot Approval Group. Sensemaker asked the BAG to require an opt-out mechanism, lamenting that Hagerman could 'force something upon people who expressly ask to be excluded'. Many more users who had previously left their comments unsigned or signed with non-standard signatures also began to make themselves visible.

In the ensuing discussion – which was comprised of BAG members, administrators, and other Wikipedians – it became clear that this was not simply a debate about signatures and timestamps. The debate had become a full-blown controversy about the morality of delegating social tasks to technologies, and it seemed that most of the participants were aware that they had entered a new territory. There had been debates about bots in Wikipedia before, but most were not about bots *per se*, instead revolving around whether a particular task – which just happened to be performed by a bot – was a good idea or not. If there was a consensus for performing the task, the bot was approved and began operating; if there was no consensus, the bot was rejected, or suspended if it had already been operating. In the case of HagermanBot, critics increasingly began to claim that there was something fundamentally different between humans sporadically correcting violations of a generally-accepted norm and a bot relentlessly ensuring total compliance with its interpretation of this norm. For them, the burden was on Hagerman and his allies to reach a consensus in favor of the current implementation of the bot if they wanted to keep it operating.

The bot's supporters rejected this, claiming that HagermanBot was only acting in line with a well-established and agreed-upon understanding that the community had reached regarding the importance of signatures in discussion spaces. For them, the burden was on the critics to reach a consensus to amend the Signatures guideline if they wanted to stop the bot from operating. Hagerman portrayed the two supported opt-out systems (!NOSIGN! and <!--Disable HagermanBot-->) not as ways for users to decide for themselves if they ought to abide by the Signatures guideline, but rather to keep the bot from signing particular contributions to talk pages that are not actually comments and therefore, according to the guideline, do not need to be signed. These would include the various informational banners routinely placed on talk pages to let editors know, for example, that the article is being proposed for deletion or that it will be featured on the main page the next week. From a design standpoint, HagermanBot thus assumed total editorial compliance with the Signatures guideline: the two opt-out features were to ensure more conformity, not less, by allowing users to tell the bot when a Signature would be unwarranted according to the guideline. Users who were opposed to the Signatures guideline in general could use the tedious feature to prevent the bot from enforcing the guideline when they made comments, but Hagerman begged them not to opt-out in this manner.

HagermanBot's allies were thus able specifically to articulate a shared vision of how discussion spaces were and ought to be, placing strong moral emphasis on the role of signatures and timestamps in maintaining discursive order and furthering the ideals of openness and verifiability. Like all approved bots that came before it, HagermanBot was acting to realize a community-sanctioned vision of what Wikipedia was and how it ought to be. The Signatures guideline was clear, stating that users were not to be punished for failing to sign their comments, but that all signatures should be signed, given that signatures were essential to the smooth operation of Wikipedia as an open, discussion-based community.

Yet this proved inadequate to settle the controversy, because those opposed to HagermanBot were articulating a different view of Wikipedia – one that did not directly contest the claims made regarding the importance of signatures, discussion pages, and communicative conventions. Instead, those like Sensemaker advanced an opposing view of how users, and especially bot operators, ought to act toward each other in Wikipedia, a view that drew heavily on notions of mutual respect:

> Concerning your emphasis on the advantages of the bot I am sure that it might be somewhat convenient for you or others to use this bot to sign everything I write. However, I have now specifically requested to not have it implemented against my will. I would not force something upon you that you expressly said you did not want for my convenience. Now I humbly request that the same basic courtesy be extended to me. -Sensemaker

For HagermanBot's allies, these objections were categorically interpreted as irrational, malicious, or indicative of what Rich Farmbrough called 'botophobia'. While this seems to be a pejorative description that would strengthen Hagerman's position, it restructured the controversy and allowed it to be settled in Sensemaker's favor. In entering the debate, Farmbrough argued that while Hagerman and his allies were entirely correct in their interpretation of the Signatures guideline, Hagerman should still allow an opt-out system:

> On the one hand, you can sign your edits (or not) how you like, on the other it is quite acceptable for another user to add either the userid, time or both to a talk edit which doesn't conatin them. Nonetheless it might be worth allowing users to opt out of an automatic system - with an opt out list on a WP page (the technical details will be obvious to you)- after all everything is in history. This is part of the 'bots are better behaved than people' mentality whihc is needed to avoid botophobia. *Rich Farmbrough*, 18:22 6 December 2006 (GMT).

Such a mediation between incommensurable views was sufficient to resolve the compromise. Declarations of either side's entitlements, largely articulated in the language of positive rights, were displaced by the notion of responsibility, good behavior, and mutual respect. What it meant to be a good bot operator now included maintaining good relations with editors who objected to bots or else risk a wave of anti-bot sentiment. The next day Hagerman agreed, and the issue was settled.

**An Unexpected Ally**
While the opt-out list may seem like a concession made by Hagerman, it proved to be one of his strongest allies in defending HagermanBot from detractors, who were arriving in numbers to his user talk page and other spaces, even after the Sensemaker/Hagerman dispute had been settled. Most users left value-neutral bug reports or positive expressions of gratitude, but a small but steadily-increasing number of editors continued to complain about the bot's automatic signing of their comments. The arguments made against HagermanBot were diverse in their rationales, ranging from complaints based on annoyance to accusations that the bot violated long-established rights of editors in Wikipedia. As one editor asked:

> *Who gave you the right to do this?*
> It is not mandatory that we sign, AFAIK. Instead of concocting this silly hack, why not get the official policy changed? I suppose you effectively did that by getting permission to run your bot on WP. How did you manage that anyway? (I won't bother with typing the fourtildas).
>
>> It isn't a policy, however, it is a guideline. You can view its approval at Wikipedia:Bots/Requests for approval/HagermanBot. Feel free to opt out if you don't want to use it. Best, Hagerman(talk) 02:29, 5 January 2007 (UTC)

As seen in Hagerman's reply to this objection, a few human allies were helpful in rebutting the objections made against his bot: the members of the Bot Approval Group, who had reviewed and approved the bot according to established protocols. The Signatures guideline – including the distinction between guidelines and policies – was also invoked to justify HagermanBot's actions, as shown in both examples. It would seem that these actors, who were generally taken to draw their legitimacy from a broad, project-wide consensus, would have been the most powerful allies that Hagerman could deploy in support of HagermanBot's actions and its vision of how discussion spaces in Wikipedia ought to operate. However, a much stronger ally proved to be the opt-out list through which angry editors could be made to lose interest in the debate altogether. It is this last actor that was most widely used by Hagerman and his human allies, who began to routinely use the opt-out list to respond to a wide array of objections made against the bot.

The strength of the opt-out list was its flexibility in rebutting the objections from two kinds of arguments: first, the largely under-articulated claims that the bot was annoying or troublesome to them; and second, the ideological or rights-based arguments that the bot was acting against fundamental principles of the project's normative structure. The first argument was easy to rebut, given that the opt-out list completely responded to their more practical concerns. In contrast, those making the second kind of argument called forth juridico-political concepts of rights, autonomy, and freedom. Yet the same opt-out list could be invoked in HagermanBot's defense against these actors, as it foreclosed their individual claims that the bot was violating their editorial rights. While objectors would have preferred that the bot use an opt-in list to preemptively ensure the rights of all editors, the opt-out list allowed HagermanBot to be characterized as a supremely respectful entity that was, as the new philosophy of bot building held, 'better behaved than people'.

**Exclusion Compliance**
HagermanBot's two new features – the opt-out list and the <!--Disable HagermanBot--> tag – soon became regular players in Wikipedia, especially among the bot development community. Rich Farmbrough saw the value of these non-human actors who helped settle the HagermanBot controversy and wanted to extend such functionality to other bots; however, its idiosyncratic mechanisms were unwieldy. About a week after HagermanBot implemented the opt-out list, he was involved in a discussion about a proposed bot named PocKleanBot, which was described by its operator PockingtonDan as a 'nag-bot' that would leave messages for users on their talk pages if articles they had edited were flagged for cleanup. It was unleashed without approval by the BAG and was promptly banned; in the ensuing discussion, many editors and administrators called for the 'spam bot' to be opt-in only. However, PockingtonDan argued that the bot would not be useful without sending unsolicited messages. In response, Rich Farmbrough suggested the same opt-out solution that had settled the HagermanBot controversy. However, seeing a need for extending this functionality to all possible bots, he created a template called {{nobots}}, which was to perform the same function as HagermanBot's exclusion tag, except apply to all compliant bots.

Most templates contain a pre-written message, but the message attached to the nobots template was blank, thus it would not change the page for viewers but could be added by editors and detected by bots that downloaded its source code. If a user placed the text {{nobots}} on their user page, any bot that supported the standard would not edit that page in any fashion. A user could also allow only specific bots access by writing, for example, {{nobotslallow=HagermanBot}}. In short, {{nobots}} was a sign that users could place on pages to signal to certain bots that they were either welcome or not welcome to edit on that page, with no actual technical ability to restrict non-compliant bots from editing. A bot would have to be built such that it looked for this template and respected it; in the case of PockingtonBot, incorporating this feature was required by the BAG in order to approve the bot.

While the controversy of PocKleanBot was settled by PockingtonDan bowing to the pressure of the BAG and removing it from operation, the template fared much better in the bot development community. Along with Farmbrough, Hagerman was one of the key actors in developing the initial specification for {{nobots}}, along with Ram-Man, a member of the Bot Approval Group. On 18 December, Hagerman announced that HagermanBot was now 'nobots aware' on the template's talk page, the first recorded bot to become what would later be called exclusion compliant – a term that Hagerman crafted. After some confusion with semantics, the template was copied to {{bots}} and remained relatively stable for the next few months as it gained acceptance and increasing use among bots. After HagermanBot, the next bot to be made exclusion-compliant was AzaBot, created to leave user talk page messages for users in a certain specialized discussion after an outcome was reached. AzaToth submitted the proposal to the BAG on 20 December, which was approved by Ram-Man that same day. In his decision, Ram-Man asked AzaToth to make the bot comply with {{bots}}, implementing an opt-out mechanism to 'respect their wishes'. Ram-Man also asked for AzaToth to share the source code that made this mechanism possible.

AzaToth quickly wrote a seventy-five line function in the programming language Python that incorporated compliance with this new standard, publishing it to the bot development community. This soon became fine-tuned and reduced to a four-line snippet of code, ported to five different programming languages such that nearly any bot operator could copy and paste it into their bot's code to achieve exclusion compliance. As members of the bot development community created software frameworks to facilitate bot programming, this code was eventually incorporated and enabled by default. Through the efforts of those in the BAG and the bot operator community – especially Farmborough, Hagerman, and Ram-Man – exclusion compliance became a requirement for many bots, implemented first to settle existing controversies and eventually becoming a pre-emptive mechanism for inhibiting conflict between bot editors and the community. While it was never mandatory, many bot operators had to argue why their bot should not be required to implement such features upon review by the BAG, and failure to implement exclusion compliance or opt-out lists soon became nonnegotiable grounds for denying some bot requests.

Debates about newsletter delivery bots – which exploded in popularity as the various editorial subcommunities organized in 2007 – became a site of articulation regarding this issue. Many bots were proposed that would automatically deliver a group's newsletter or targeted message to all its members. When the first of these bots began operating, conflicts initially emerged between editors who felt they had received unsolicted spam and bot operators who thought they were providing a valuable service. Opt-out mechanisms were used to settle these disputes, although in many cases the bots already incorporated such features but did not make them visible to recipients. In response, a set of informal criteria was soon formed by members of the BAG to ease these proposals. One requirement was implementation of some opt-out mechanism, either via exclusion compliance or an opt-out list; another was including information about opting-out in each newsletter delivery. Such requirements settled many controversies between editors and bot operators, and soon, bot approval policies were updated to officially indicate that no newsletter bots would be approved by the BAG until they were proven to sufficiently respect the wishes of editors who did not want interference from such bots.

**Conclusion**
The case of HagermanBot shows us how a weak but pre-existing social norm was controversially reified into a technological actor. Yet there is also a more nuanced dynamic between human and non-humans at play, as this controversy regarding the delegation of work to bots was settled by constructing a new set of technical and social artifacts – artifacts that the Wikipedian bot development community used in future debates. HagermanBot complicates accounts of the project's order that rely almost exclusively on social artifacts, showing that these non-human editors have a significant effect on how the project's norms are enforced. While much human work is performed in settling controversies, the bot development process can be a moment of articulation and contestation for what were previously taken to be uncontroversial expectations.

At the most basic level, there are many organizational restrictions on bot development, such as policies, guidelines, and a committee that must approve all bots before operation. Yet bots are also limited by their own power; in universally and uniformly acting to realize a particular normatively-charged vision of how articles ought to look or how editors ought to act, they often act rashly and make certain unstated assumptions quite visible. With HagermanBot, instantly signing the unsigned comments left by every editor brought to light differences in how two previously invisible groups interpreted a vague guideline. This is because, like Bruno Latour's speed bumps, bots are ruthlessly moral; just as a speed bump will punish both reckless drivers and ambulances in its quest to maintain order on roads, so will bots often take a particular view of Wikipedia to its logical extreme. This makes it difficult to think of bot operators as power users who silently deploy bots to further increase their power in the community.

The case of HagermanBot further illustrates that the negotiation of a bot's source code is not a purely normative affair in which participants discuss the kind of editorial environment that is to be enforced by such an actor. Following Latour, the HagermanBot controversy shows that these articulations can be both material and semiotic, that is, with intentions being expressed both in technologies and discourse, and such meanings are mutually interdependent. HagermanBot's opt-out mechanisms, for example, experienced a dramatic reversal, having first been articulated to ensure that the bot only signed edits that were actually comments – not a way for rogue editors to abandon the guideline at their whim. Yet within a new understanding of how bots and bot operators ought to act within the Wikipedian community, this translated into a way of showing respect for dissenters, with a new opt-out mechanism created to stave off 'botophobia'.

What is most notable about the HagermanBot controversy is that it marks a turning point in the understanding of what kinds of worldviews bots work to realize. Prior to HagermanBot, Wikipedian bot operation could be said to take place in a weakly technologically determinist mode, in which bots reified a vision of how the world of Wikipedia ought to be, once that vision was agreed upon by the community. Post-HagermanBot and with the rise of exclusion compliance, certain technical features of bots articulated a vision of how bots and their operators ought to relate to the community. In fact, this material-semiotic chain of meaning repeatedly oscillated between technical and discursive articulations. This persistent notion that 'bots are better behaved than people', which Hagerman articulated in the form of the opt-out mechanism, became standardized in a semiotic marker: Rich Farmborough's {{bots}} template. Compliance with this template was articulated in AzaToth's software code, which was translated into a number of programming languages such that any bot operator could easily make their bot articulate this notion of respect. Passing back into the semiotic, including this code gained the moniker of 'exclusion compliant', and this condition became regularly incorporated into BAG bot approval discussions.

In all, bots defy simple single-sided categorizations: they are both editors and software, social and technical, discursive and material, as well as assembled and autonomous. One-sided determinisms and constructionisms, while tempting, are insufficient to fully explain the complicated ways in which these bots have become vital members of the Wikipedian community.
In understanding the relationship that bots have to the world around them, we must trace how bots come to articulate and be articulated within a heterogeneous assemblage. Only then can we realize that the question of who or what is in control of Wikipedia is far less interesting than the question of how control operates across a diverse and multi-faceted sociotechnical environment.